\input amssym.tex
\input epsf
\epsfclipon

% Page layout

\magnification=\magstephalf
\hsize=14.0 true cm
\vsize=19 true cm
\hoffset=1.0 true cm
\voffset=2.0 true cm

\abovedisplayskip=12pt plus 3pt minus 3pt
\belowdisplayskip=12pt plus 3pt minus 3pt
\parindent=1.0em

% Fonts

\font\sixrm=cmr6
\font\eightrm=cmr8
\font\ninerm=cmr9

\font\sixi=cmmi6
\font\eighti=cmmi8
\font\ninei=cmmi9

\font\sixsy=cmsy6
\font\eightsy=cmsy8
\font\ninesy=cmsy9

\font\sixbf=cmbx6
\font\eightbf=cmbx8
\font\ninebf=cmbx9

\font\eightit=cmti8
\font\nineit=cmti9

\font\eightsl=cmsl8
\font\ninesl=cmsl9

\font\sixss=cmss8 at 8 true pt
\font\sevenss=cmss9 at 9 true pt
\font\eightss=cmss8
\font\niness=cmss9
\font\tenss=cmss10

 at 12 true pt
 at 12 true pt
\font\bigrm=cmr10 at 12 true pt
 at 12 true pt
 at 12 true pt

 at 16 true pt
%\font\Bigsy=cmsy12 at 16 true pt
%\font\Bigex=cmex12 at 16 true pt
 at 16 true pt
\font\Bigrm=cmr12 at 16 true pt
 at 16 true pt
 at 16 true pt

\catcode`@=11
\newfam\ssfam

\def\tenpoint{\def\rm{\fam0\tenrm}%
    \textfont0=\tenrm \scriptfont0=\sevenrm \scriptscriptfont0=\fiverm
    \textfont1=\teni  \scriptfont1=\seveni  \scriptscriptfont1=\fivei
    \textfont2=\tensy \scriptfont2=\sevensy \scriptscriptfont2=\fivesy
    \textfont3=\tenex \scriptfont3=\tenex   \scriptscriptfont3=\tenex
    \textfont\itfam=\tenit                  \def\it{\fam\itfam\tenit}%
    \textfont\slfam=\tensl                  \def\sl{\fam\slfam\tensl}%
    \textfont\bffam=\tenbf \scriptfont\bffam=\sevenbf
    \scriptscriptfont\bffam=\fivebf
                                            \def\bf{\fam\bffam\tenbf}%
    \textfont\ssfam=\tenss \scriptfont\ssfam=\sevenss
    \scriptscriptfont\ssfam=\sevenss
                                            \def\ss{\fam\ssfam\tenss}%
    \normalbaselineskip=13pt
    \setbox\strutbox=\hbox{\vrule height8.5pt depth3.5pt width0pt}%
    \let\big=\tenbig
    \normalbaselines\rm}

\def\ninepoint{\def\rm{\fam0\ninerm}%
    \textfont0=\ninerm      \scriptfont0=\sixrm
                            \scriptscriptfont0=\fiverm
    \textfont1=\ninei       \scriptfont1=\sixi
                            \scriptscriptfont1=\fivei
    \textfont2=\ninesy      \scriptfont2=\sixsy
                            \scriptscriptfont2=\fivesy
    \textfont3=\tenex       \scriptfont3=\tenex
                            \scriptscriptfont3=\tenex
    \textfont\itfam=\nineit \def\it{\fam\itfam\nineit}%
    \textfont\slfam=\ninesl \def\sl{\fam\slfam\ninesl}%
    \textfont\bffam=\ninebf \scriptfont\bffam=\sixbf
                            \scriptscriptfont\bffam=\fivebf
                            \def\bf{\fam\bffam\ninebf}%
    \textfont\ssfam=\niness \scriptfont\ssfam=\sixss
                            \scriptscriptfont\ssfam=\sixss
                            \def\ss{\fam\ssfam\niness}%
    \normalbaselineskip=12pt
    \setbox\strutbox=\hbox{\vrule height8.0pt depth3.0pt width0pt}%
    \let\big=\ninebig
    \normalbaselines\rm}

\def\eightpoint{\def\rm{\fam0\eightrm}%
    \textfont0=\eightrm      \scriptfont0=\sixrm
                             \scriptscriptfont0=\fiverm
    \textfont1=\eighti       \scriptfont1=\sixi
                             \scriptscriptfont1=\fivei
    \textfont2=\eightsy      \scriptfont2=\sixsy
                             \scriptscriptfont2=\fivesy
    \textfont3=\tenex        \scriptfont3=\tenex
                             \scriptscriptfont3=\tenex
    \textfont\itfam=\eightit \def\it{\fam\itfam\eightit}%
    \textfont\slfam=\eightsl \def\sl{\fam\slfam\eightsl}%
    \textfont\bffam=\eightbf \scriptfont\bffam=\sixbf
                             \scriptscriptfont\bffam=\fivebf
                             \def\bf{\fam\bffam\eightbf}%
    \textfont\ssfam=\eightss \scriptfont\ssfam=\sixss
                             \scriptscriptfont\ssfam=\sixss
                             \def\ss{\fam\ssfam\eightss}%
    \normalbaselineskip=10pt
    \setbox\strutbox=\hbox{\vrule height7.0pt depth2.0pt width0pt}%
    \let\big=\eightbig
    \normalbaselines\rm}

\def\tenbig#1{{\hbox{$\left#1\vbox to8.5pt{}\right.\n@space$}}}
\def\ninebig#1{{\hbox{$\textfont0=\tenrm\textfont2=\tensy
                       \left#1\vbox to7.25pt{}\right.\n@space$}}}
\def\eightbig#1{{\hbox{$\textfont0=\ninerm\textfont2=\ninesy
                       \left#1\vbox to6.5pt{}\right.\n@space$}}}

\font\sectionfont=cmbx10
\font\subsectionfont=cmti10

\def\figurecaptionfont{\ninepoint}
\def\tablecaptionfont{\ninepoint}
\def\footnotefont{\eightpoint}

% New count registers

\newcount\equationno
\newcount\bibitemno
\newcount\figureno
\newcount\tableno

\equationno=0
\bibitemno=0
\figureno=0
\tableno=0
%\advance\pageno by -1

% Footline

\footline={\ifnum\pageno=0{\hfil}\else
{\hss\rm\the\pageno\hss}\fi}

% Section macro

\def\section #1. #2 \par
{\vskip0pt plus .10\vsize\penalty-100 \vskip0pt plus-.10\vsize
\vskip 1.6 true cm plus 0.2 true cm minus 0.2 true cm
\global\def\equationlabel{#1}
\global\equationno=0
\leftline{\sectionfont #1. #2}\par
\immediate\write\terminal{Section #1. #2}
\vskip 0.7 true cm plus 0.1 true cm minus 0.1 true cm
\noindent}

% Subsection macro

\def\subsection #1 \par
{\vskip0pt plus 0.8 true cm\penalty-50 \vskip0pt plus-0.8 true cm
\vskip2.5ex plus 0.1ex minus 0.1ex
\leftline{\subsectionfont #1}\par
\immediate\write\terminal{Subsection #1}
\vskip1.0ex plus 0.1ex minus 0.1ex
\noindent}

% Appendix macro

\def\appendix #1. #2 \par
{\vskip0pt plus .20\vsize\penalty-100 \vskip0pt plus-.20\vsize
\vskip 1.6 true cm plus 0.2 true cm minus 0.2 true cm
\global\def\equationlabel{\hbox{\rm#1}}
\global\equationno=0
\leftline{\sectionfont Appendix #1. #2}\par
\immediate\write\terminal{Appendix #1. #2}
\vskip 0.7 true cm plus 0.1 true cm minus 0.1 true cm
\noindent}

%\def\appendix #1. #2 \par
%{\vskip0pt plus .20\vsize\penalty-100 \vskip0pt plus-.20\vsize
%\vskip 1.6 true cm plus 0.2 true cm minus 0.2 true cm
%\global\def\equationlabel{\hbox{\rm#1}}
%\global\equationno=0
%\leftline{\sectionfont Appendix #1. #2}\par
%\immediate\write\terminal{Appendix #1. #2}
%\vskip 0.7 true cm plus 0.1 true cm minus 0.1 true cm
%\noindent}

% Displayed equations

\def\equation#1{$$\displaylines{\qquad #1}$$}
\def\enum{\global\advance\equationno by 1
\hfill\llap{{\rm(\equationlabel.\the\equationno)}}}

\def\next#1{\cr\noalign{\vskip#1}\qquad}

% Bibliography macro, references

\def\ifundefined#1{\expandafter\ifx\csname#1\endcsname\relax}

\def\ref#1{\ifundefined{#1}?\immediate\write\terminal{unknown reference
on page \the\pageno}\else\csname#1\endcsname\fi}

\newwrite\terminal
\newwrite\bibitemlist

\def\bibitem#1#2\par{\global\advance\bibitemno by 1
\immediate\write\bibitemlist{\string\def
\expandafter\string\csname#1\endcsname
{\the\bibitemno}}
\item{[\the\bibitemno]}#2\par}

\def\beginbibliography{
\vskip0pt plus .15\vsize\penalty-100 \vskip0pt plus-.15\vsize
\vskip 1.2 true cm plus 0.2 true cm minus 0.2 true cm
\leftline{\sectionfont References}\par
\immediate\write\terminal{References}
\immediate\openout\bibitemlist=biblist
\frenchspacing\parindent=1.8em
\vskip 0.5 true cm plus 0.1 true cm minus 0.1 true cm}

\def\endbibliography{
\immediate\closeout\bibitemlist
\nonfrenchspacing\parindent=1.0em}

% Figure and table captions

\def\figurecaption#1{\global\advance\figureno by 1
\narrower\figurecaptionfont
Fig.~\the\figureno. #1}

\def\tablecaption#1{\global\advance\tableno by 1
\vbox to 0.5 true cm { }
\centerline{\tablecaptionfont%
Table~\the\tableno. #1}
\vskip-0.4 true cm}

\def\thicktablerule{\hrule height1pt}
\def\thintablerule{\hrule height0.4pt}

\tenpoint

\immediate\openin\bibitemlist=biblist
\ifeof\bibitemlist\immediate\closein\bibitemlist
\else\immediate\closein\bibitemlist
\input biblist \fi

% current year and month

\def\thismonth{\ifcase\month\or
January\or February\or March\or April\or May\or June\or
July\or August\or September\or October\or November\or December\fi}

% Definitions and abbreviations

% Roman letters in math formulae

\def\rme{{\rm e}}

\def\rmo{{\rm o}}

% Real and integer numbers

% Special relations and symbols

\def\proof{\noindent{\sl Proof:}\kern0.6em}

\def\frac#1#2{\hbox{$#1\over#2$}}
\def\dual{\mathstrut^*\kern-0.1em}

\def\lvec#1{\setbox0=\hbox{$#1$}
    \setbox1=\hbox{$\scriptstyle\leftarrow$}
    #1\kern-\wd0\smash{
    \raise\ht0\hbox{$\raise1pt\hbox{$\scriptstyle\leftarrow$}$}}
    \kern-\wd1\kern\wd0}
\def\rvec#1{\setbox0=\hbox{$#1$}
    \setbox1=\hbox{$\scriptstyle\rightarrow$}
    #1\kern-\wd0\smash{
    \raise\ht0\hbox{$\raise1pt\hbox{$\scriptstyle\rightarrow$}$}}
    \kern-\wd1\kern\wd0}
\def\slash#1{\setbox0=\hbox{$#1$}\setbox1=\hbox{$\kern1pt/$}
    #1\kern-\wd0\kern1pt/\kern-\wd1\kern\wd0}

% Lattice derivatives

\def\nab#1{{\nabla_{#1}}}
\def\nabstar#1{{\nabla\kern0.5pt\smash{\raise 4.5pt\hbox{$\ast$}}
               \kern-5.5pt_{#1}}}

\def\drvstar#1{{\partial\kern0.5pt\smash{\raise 4.5pt\hbox{$\ast$}}
               \kern-6.0pt_{#1}}}

\def\ldrvstar#1{{\lvec{\,\partial}\kern-0.5pt\smash{\raise 4.5pt\hbox{$\ast$}}
               \kern-5.0pt_{#1}}}

% Units

\def\MeV{{\rm MeV}}

\def\fm{{\rm fm}}

% Constants

% Fields

% Dirac matrices

\def\dirac#1{\gamma_{#1}}
\def\diracstar#1#2{
    \setbox0=\hbox{$\gamma$}\setbox1=\hbox{$\gamma_{#1}$}
    \gamma_{#1}\kern-\wd1\kern\wd0
    \smash{\raise4.5pt\hbox{$\scriptstyle#2$}}}

% Gauge group

\def\Ad{{\rm Ad}\kern0.1em}

% Dirac operator

\def\Dw{D_{\rm w}}
\def\Dhat{\hat{D}}
\def\Msap{M_{\rm sap}}

% Blocks and block operators

\def\B{\Lambda}
\def\Bs{\B^{\kern-0.1em*}}
\def\dB{\partial\B}
\def\dBs{\partial\Bs}
\def\DB{D_{\B}}
\def\DBs{D_{\Bs_{\vphantom{1}}}}
\def\DdB{D_{\dB}}
\def\DdBs{D_{\dBs_{\vphantom{1}}}}

\def\Om{\Omega}
\def\Oms{\Omega^{\ast}}
\def\dOm{\partial\Om}
\def\dOms{\partial\Oms}
\def\DOm{D_{\Om}}
\def\DOms{D_{\Oms_{\vphantom{1}}}}
\def\DdOm{D_{\dOm}}
\def\DdOms{D_{\dOms_{\vphantom{1}}}}

% Schwarz procedure

\def\ncy{n_{\rm cy}}
\def\nmr{n_{\rm mr}}
\def\nkv{n_{\rm kv}}

% Quark masses

\def\mq{m_q}
\def\mc{m_c}

\vbox{\vskip0.0cm}
\rightline{CERN-TH/2003-250}

\vskip 1.6cm 
\centerline{\Bigrm Solution of the Dirac equation in lattice QCD}
\vskip 0.3 true cm
\centerline{\Bigrm using a domain decomposition method}
\vskip 0.6 true cm
\centerline{\bigrm Martin L\"uscher}
\vskip1ex
\centerline{\it CERN, Theory Division}
\centerline{\it CH-1211 Geneva 23, Switzerland}
\vskip 0.8 true cm
\thintablerule
\vskip 2.0ex
\ninepoint
\leftline{\bf Abstract}
\vskip 1.0ex\noindent
Efficient algorithms for the solution of
partial differential equations on parallel computers
are often based on domain decomposition methods.
Schwarz preconditioners combined with standard Krylov 
space solvers are widely used in this context, and such
a combination is shown here to perform very well in the case
of the Wilson--Dirac equation in lattice QCD.
In particular, with respect to even-odd preconditioned
solvers, the communication overhead is significantly reduced,
which allows 
the computational work to be distributed over a large number 
of processors with only small parallelization losses.
\vskip 2.0ex
\thintablerule

\tenpoint

%\vskip-0.3cm

\section 1. Introduction

At present, numerical simulations of lattice QCD 
including sea-quark effects are still limited to
relatively small lattices and large quark masses.
The rapid increase in the computer power that is available
for these calculations will certainly help to improve the situation,
but it may also be necessary to develop better algorithms
to be able to reach the chiral regime, where the 
effects associated with the spontaneous breaking of 
chiral symmetry become important.

In this connection the applicability of  
domain decomposition methods [\ref{QuarteroniValli},\ref{Saad}]
seems worth studying, given the fact that
the standard formulations of lattice QCD involve nearest-neighbour
interactions only. A first step in this direction was
made in ref.~[\ref{LuscherSchwarz}], where 
a simulation algorithm for two-flavour lattice QCD was proposed,
which can be regarded as an implementation of 
the classic Schwarz alternating procedure [\ref{Schwarz}] 
at the quantum level. 

The classic Schwarz procedure
may also be used directly for the solution of 
the lattice Dirac equation,
and the aim in the present paper is to show that this leads to 
a competitive solver, particularly so if the computational
work is to be distributed over a large number of 
nodes of a parallel computer.
While this result is immediately relevant to the QCD
simulation algorithms that are currently in use, the 
study of the Schwarz procedure in this context
also serves as a preparation for 
the implementation of the proposed Schwarz simulation algorithm
[\ref{LuscherSchwarz}].

For simplicity only the standard Wilson formulation of lattice QCD
[\ref{Wilson}] will be considered, 
but O($a$) improvement [\ref{SW},\ref{OaImp}]
can easily be included and the basic strategies are in any case 
expected to be more generally applicable.
The Schwarz procedure for the solution 
of the lattice Dirac equation looks fairly natural in this framework
and is explained in all detail.
By itself this method is not very efficient,
but it can be combined, 
in the form of a preconditioner for the Wilson--Dirac operator,
with the GCR Krylov space accelerator.
The excellent performance of
the resulting algorithm is then demonstrated
by a series of test runs
on a recent PC-cluster with $64$ processors.

\section 2. Schwarz alternating procedure

\vskip-2.5ex

\subsection 2.1 Lattice Dirac operator

As usual the fields in lattice QCD are assumed to live
on the sites $x$ of a hypercubic four-dimensional
lattice of finite extent.
Periodic boundary conditions are imposed in all directions
and the lattice spacing is set to unity for notational convenience.

On the lattice 
the SU(3) gauge field is represented by group-valued 
link variables $U(x,\mu)$, $\mu=0,\ldots,3$, while 
the quark fields $\psi(x)$ carry Dirac and colour indices
as in the continuum theory.
The gauge-covariant forward and backward difference operators are then
given by 
\equation{
   \nab{\mu}\psi(x)=U(x,\mu)\psi(x+\hat{\mu})-\psi(x),
   \enum
   \next{2ex}
   \nabstar{\mu}\psi(x)=\psi(x)-U(x-\hat{\mu},\mu)^{-1}\psi(x-\hat{\mu}),
   \enum
}
where $\hat{\mu}$ denotes the unit vector in direction $\mu$, 
and using these the Wilson--Dirac operator may be written in the form
\equation{
   \Dw=\frac{1}{2}\left\{\dirac{\mu}\left(\nabstar{\mu}+\nab{\mu}\right)
   -\nabstar{\mu}\nab{\mu}\right\}.
   \enum
}
A sum over $\mu$ is implied here
and the Dirac matrices $\dirac{\mu}$ are taken to be hermitian.
In the following, the Schwarz procedure will be set up
for the linear equation
\equation{
   D\psi(x)=\eta(x),
   \qquad D\equiv \Dw+m_0,
   \enum
}
in which $\eta$ is an arbitrary source field and $m_0$ the bare quark mass.

\subsection 2.2 Block terminology

Like any other domain decomposition method, the Schwarz procedure
operates on a covering of the lattice by 
a collection of domains.
In the present context
it is natural to choose the domains to be
rectangular blocks of lattice points.
Any such block is completely
characterized by the lengths of its edges and by its position.
In practice the blocks are usually taken to be fairly small,
particularly on parallel computers, where
for reasons of efficiency they should be contained
in the local lattices.\kern1pt\footnote{$\dagger$}{\footnotefont%
Parallel programs in lattice QCD are usually
based on a division of the lattice
into sublattices that are associated to the processors of the computer
or to different program threads.
The term ``local lattice" refers to these sublattices.}

\topinsert
\vbox{
\vskip0.0cm
\epsfxsize=3.0cm\hskip4.5cm\epsfbox{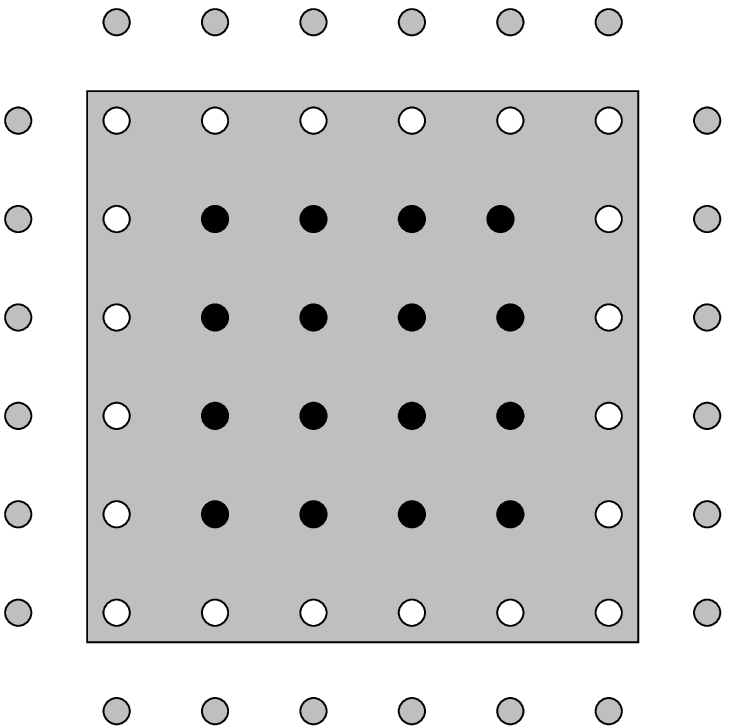}
\vskip0.2cm
\figurecaption{%
Slice of 
a $6^4$ block $\B$ (points inside the grey square) and its
exterior boundary $\dB$ (points outside the square)
along a two-dimensional lattice plane.
The white points in the square represent 
the interior boundary $\dBs$ of the block.
}
\vskip0.0cm
}
\endinsert

The exterior boundary $\dB$ of a given block $\B$ is defined to 
be the set of all lattice points that have distance $1$ from $\B$ (see
fig.~1). 
Each exterior boundary point has a closest ``partner" point in the block.
The interior boundary $\dBs$ of $\B$ consists of all these points,
and the set of points that are not in the block is denoted by
$\Bs$.

In position space the lattice Dirac operator is a sparse matrix
that assumes the block-diagonal form
\equation{
  D=\pmatrix{\DB\hfill\kern-0.5em &\DdB\hfill \cr
             \noalign{\vskip1ex}
             \DdBs\hfill\kern-0.5em &\DBs\hfill \cr}
  \enum
}
if the lattice points are ordered so that those in $\B$ come first.
The operator $\DB$, for example, acts on quark fields on $\B$
in precisely the same way as $D$, except that all terms
involving the
exterior boundary points are set to zero (which is equivalent to
imposing Dirichlet boundary conditions on $\dB$).

It is often convenient to consider
$\DB$, $\DBs$, $\DdB$ and $\DdBs$ to operate on
quark fields on the whole lattice
rather than on fields that are defined on $\B$ or $\Bs$ only.
The embedding is done in the obvious way by
padding with zeros so that eq.~(2.5), for example, may equivalently
be written in the form 
\equation{
   D=\DB+\DBs+\DdB+\DdBs.
   \enum
}
This notation is perhaps slightly abusive
but it will usually be clear from the context 
which interpretation applies.

\subsection 2.3 Definition of the Schwarz procedure

In the form in which it was originally conceived [\ref{Schwarz}],
the Schwarz procedure is assumed to operate on overlapping domains.
The convergence of the method actually tends to be better the larger
the overlaps are [\ref{QuarteroniValli}].
On the other hand, more domains are then
required to cover the lattice, which slows down the computations,
particularly in four dimensions, where the average occupation number
is rapidly growing with the overlap size.
If the Schwarz procedure is only used as a preconditioner
(as will be the case here), it can thus be advantageous to choose 
the domains to be non-overlapping.

\topinsert
\vbox{
\vskip0.0cm
\epsfxsize=7.0cm\hskip2.5cm\epsfbox{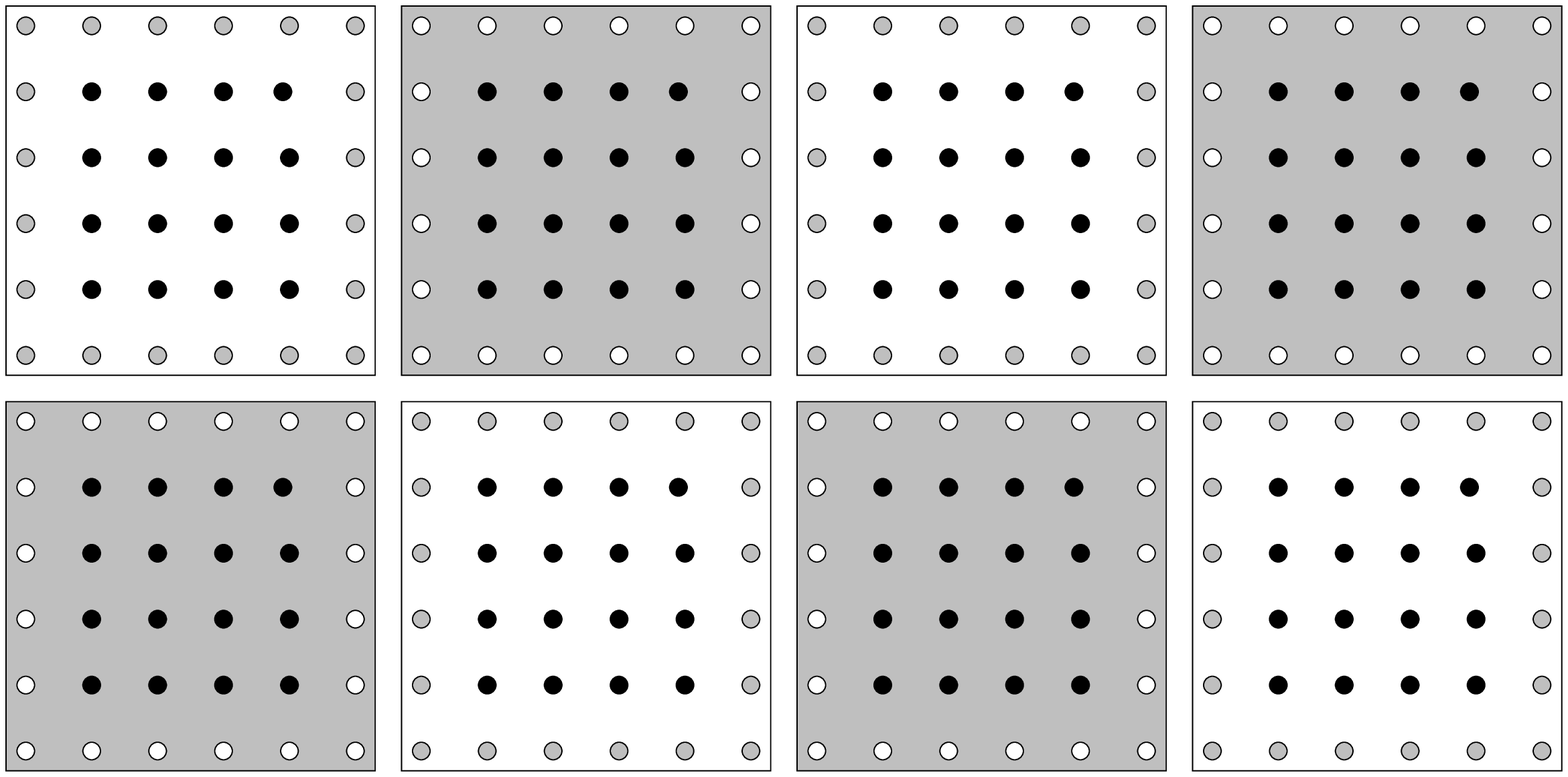}
\vskip0.3cm
\figurecaption{%
Two-dimensional cross-section of a grid of non-overlapping $6^4$ blocks
that covers a lattice of size $24\times12^3$. 
Such grids can be chessboard-coloured
if there is an even number of blocks in all dimensions.
}
\vskip0.0cm
}
\endinsert

The obvious choice is then to cover the lattice
by a regular grid of non-overlapping rectangular blocks 
of equal size such as the one shown in fig.~2.
Although this is not required for the Schwarz procedure,
it will be assumed, for technical reasons, that the block
sizes are even and that the grid can be chessboard-coloured.

The so-called multiplicative Schwarz procedure now visits
the blocks in the chosen block grid 
sequentially and updates the current approximation
$\psi$ to the solution of the Wilson--Dirac equation (2.4) there.
At the start of the iteration one can take $\psi=0$, for example.
When the algorithm arrives at block $\B$, the updated field
$\psi'$ is obtained by solving the equations
\equation{
  \left.D\psi'(x)\right|_{x\in\B}=\eta(x),
  \qquad
  \left.\psi'(x)\right|_{x\notin\B}=\psi(x).
  \enum
}
It is straightforward to check that this amounts to setting 
\equation{
  \psi'=\psi+\DB^{-1}(\eta-D\psi),
  \enum
}
where it is understood that $\DB^{-1}$ acts
on the restriction of the current residue $\eta-D\psi$ to the block $\B$.
This form of the iteration is numerically stable and is the one that
is actually programmed.

\subsection 2.4 Schwarz iteration operator

The structure of the Schwarz procedure becomes more transparent if
it is rewritten in a slightly more compact form.
To this end first note that the residue of the new approximation
(2.8) coincides with the old residue, except on the block $\B$
and on the neighbouring blocks.
So if all black and all white blocks are visited alternatingly,
it is clear that
the order of the blocks with the same colour is irrelevant and
that they can in fact be updated simultaneously. 

Effectively the algorithm
thus operates on only two domains, one being
the union $\Om$ of all black blocks and the other the union $\Oms$
of all white blocks.
With respect to this division of the lattice, 
the Dirac operator decomposes into a sum of operators
$\DOm$, $\DOms$, $\DdOm$ and $\DdOms$, as in the case
of the division defined by a block $\B$ and 
its complement $\Bs$. Some algebra then shows that 
a complete Schwarz cycle, where the quark field
is updated on all the black blocks followed by all the white
blocks, amounts to the transformation
\equation{
  \psi\to(1-KD)\psi+K\eta,
  \enum
  \next{2ex}
  K\equiv\DOm^{-1}+\DOms^{-1}-\DOms^{-1}\DdOms\DOm^{-1}.
  \enum
}
The field thus evolves from $\psi=0$ to
\equation{
  \psi=K\sum_{\nu=0}^{\ncy-1}(1-DK)^{\nu}\eta
  \enum
}
after $\ncy$ cycles,
which shows that the Schwarz procedure attempts to obtain the
solution in the form of a Neumann series with iteration operator $1-DK$.
From this point of view, the operator $K$ plays the r\^ole of a 
preconditioner for $D$.

\section 3. Preconditioned GCR algorithm

Some numerical experimenting suggests that the spectral radius of 
the iteration operator $1-DK$ is bounded by $1$, at least as long as
$D$ is not exceptionally ill-conditioned.
The rate of convergence of the Schwarz procedure is often
disappointing, however, and it is thus advisable to combine 
the method with a Krylov space accelerator
such as the generalized conjugate residual (GCR) algorithm.

\subsection 3.1 Schwarz preconditioner

The basic idea is to 
use the Schwarz procedure as a preconditioner rather than 
as a solver, and to apply the 
GCR algorithm to the preconditioned system.
As discussed in the previous section, 
the Schwarz procedure can be seen
as an operation acting on the source field, 
which produces an approximate
solution of eq.~(2.4), for any specified number $\ncy$ of 
cycles.
If the associated linear operator is denoted by $\Msap$,
the right-preconditioned equation
\equation{
  D\Msap\phi=\eta
  \enum
}
is thus expected to be
better conditioned than the original linear system (once this
equation is solved,
the solution of the latter is obtained by setting $\psi=\Msap\phi$).

From eq.~(2.11) it follows that $\Msap$ is given explicitly by
\equation{
  \Msap=K\sum_{\nu=0}^{\ncy-1}(1-DK)^{\nu}.
  \enum
}
In practice this formula is not used, however,
and the calculation of (say) $\psi=\Msap\phi$ instead proceeds along
the lines described in subsect.~2.3, i.e.~by starting from $\psi=0$
and running through the blocks in the block grid, updating $\psi$ 
on each block, with $\phi$ now being the source.

\subsection 3.2 Outline of the GCR algorithm 

The GCR algorithm is a Krylov space solver that
is mathematically equivalent to the GMRES algorithm
(see ref.~[\ref{Saad}] for example). 
It offers some technical advantages with respect to 
the latter if the preconditioner is 
implemented approximately, as will be the case here.
In the following
a variant of the GCR algorithm is described that
was introduced in the context of
the so-called GMRESR algorithm [\ref{GMRESRI},\ref{GMRESRII}].

Starting from $\psi_0=0$,
the GCR algorithm generates
a sequence of approximate solutions $\psi_k$,
$k\geq0$, of the Wilson--Dirac equation (2.4) recursively.
The associated residues are defined by
\equation{
  \rho_k=\eta-D\psi_k,
  \enum
}
and the recursion then chooses $\psi_{k+1}$ to be the field
in the linear subspace spanned by $\psi_1,\ldots,\psi_k,\Msap\rho_k$,
which minimizes $\|\rho_{k+1}\|$.

It is important to understand that the field $\Msap\rho_k$
merely defines the direction in which the previous subspace is extended.
In particular, a precise implementation of the preconditioner is 
not really required, and its definition may even vary from 
step to step. The worst that can happen in this case is that
the convergence of the algorithm slows down, but the algorithm 
will never become unstable or otherwise invalid.

\subsection 3.3 Definition of the GCR recursion

The GCR algorithm generates
two sequences of fields,
$\chi_k$ and $\xi_k$, 
that must be kept in the memory of the computer
together with the last residue $\rho_k$.
A special feature of the chosen implementation is that
the approximate
solution $\psi_k$ itself does not take part 
in the recursion and is only constructed
at the end of the calculation. 
For reasons of efficiency
the recursion may have to be restarted 
after a certain number of steps, taking the last residue
as the new source. 
This point will be discussed again in sect.~4,
together with other implementation details.

\topinsert
\vbox{
\vskip1.0ex
\centerline{\vbox{
\hsize=8.0 true cm
{\obeylines
  $\rho_0=\eta$
  \vskip0.5ex
  {\tt for} $k=0,1,2,\ldots$ {\tt do}
  \vskip0.5ex
  \quad$\xi_k=\Msap\rho_k$
  \vskip0.5ex
  \quad$\chi_k=D\xi_k$
  \vskip0.5ex
  \quad{\tt for} $l=0,\ldots,k-1$ {\tt do}
    \vskip0.5ex
    \qquad$a_{lk}=\left(\chi_l,\chi_k\right)$
    \vskip0.5ex
    \qquad$\chi_k=\chi_k-a_{lk}\chi_l$
    \vskip0.5ex
  \quad{\tt end do}
  \vskip0.5ex
  \quad$b_k=\|\chi_k\|$
  \vskip0.5ex
  \quad$\chi_k=\chi_k/b_k$
  \vskip0.5ex
  \quad$c_k=\left(\chi_k,\rho_k\right)$
  \vskip0.5ex
  \quad$\rho_{k+1}=\rho_k-c_k\chi_k$
  \vskip0.5ex
  {\tt end do}
}
}
}
\vskip2.0ex
\figurecaption{%
Pseudo-code for the GCR recursion. On the first line in the outer
loop the Schwarz preconditioner is applied to 
the current residue $\rho_k$. For the calculation of the next
residue in the last two lines, 
the auxiliary field $\chi_k$ first needs to be constructed through 
a Gram--Schmidt orthogonalization process (inner loop).
}
\vskip0.0cm
}
\endinsert

Apart from the sequences of fields, the recursion generates
sequences of complex numbers $a_{lk}$, 
$b_k$ and $c_k$ that should also be preserved. 
An explicit description of the algorithm is provided
by the pseudo-code displayed in fig.~3.
Note that the fields $\chi_k$ 
satisfy
\equation{
   \left(\chi_i,\chi_j\right)=\delta_{ij},
   \quad
   0\leq i,j\leq k,
   \enum
}
by the time $\rho_{k+1}$ is calculated. The latter is then the 
projection of the first residue $\eta$ to the 
orthogonal complement of the space spanned by the fields
$\chi_0,\ldots,\chi_k$, which coincides with the space spanned
by the fields $D\xi_0,\ldots,D\xi_k$. In other words,
\equation{
  \rho_{k+1}=\eta-\sum_{l=0}^k\alpha_lD\xi_l
  \enum
}
for some complex coefficients $\alpha_k$ chosen such that $\|\rho_{k+1}\|$
is minimized. This shows that $\rho_{k+1}$ is indeed the residue of the 
approximate solution $\psi_{k+1}$ that was defined 
in subsect.~3.2, and the pseudo-code thus implements
the GCR algorithm correctly.

\subsection 3.4 Calculation of $\psi_{k+1}$

When it is decided to stop or restart the GCR recursion 
after $k+1$ steps, the field $\psi_{k+1}$ will have to be computed. 
From the above it is obvious that 
\equation{
  \psi_{k+1}=\sum_{l=0}^k\alpha_l\xi_l,
  \enum
}
but the coefficients $\alpha_l$ [which are the same as in eq.~(3.5)]
are only implicitly known at this point.
They can, however, be calculated by 
noting that the fields $\chi_l$ satisfy
\equation{
  D\xi_l=b_l\chi_l+\sum_{i=0}^{l-1}a_{il}\chi_i,
  \quad
  l=0,1,\ldots,k,
  \enum
}
and also
\equation{
  \rho_{k+1}=\eta-\sum_{l=0}^kc_l\chi_l.
  \enum
}
Now if $\rho_{k+1}=\eta-D\psi_{k+1}$ is evaluated
by first inserting eq.~(3.6) and then eq.~(3.7), the comparison with eq.~(3.8)
leads to the linear system
\equation{
  b_l\alpha_l+\sum_{i=l+1}^ka_{li}\alpha_i=c_l,
  \quad
  l=k,k-1,\ldots,0,
  \enum
}
which can be solved straightforwardly for the coefficients 
$\alpha_l$ by back-substitution.

\section 4. Numerical implementation of the Schwarz procedure

In this section some further details of the proposed algorithm
are discussed, leaving away any purely programming issues.
The Schwarz procedure is actually not entirely trivial to program,
and a good choice of the data 
and program structures can be important
(see appendix A for some
suggestions).

\subsection 4.1 Block solver

The first question to be addressed here is how to perform
the inversion of the local Wilson--Dirac operator $\DB$
in the update step (2.8).
Clearly an exact inversion is not possible in practice,
and is in fact not needed, since an inaccurate
implementation of the Schwarz preconditioner $\Msap$
is perfectly acceptable (cf.~subsect.~3.2).

Perhaps the simplest iterative solver that may be used in this
context is the mi\-ni\-mal residual algorithm 
(ref.~[\ref{Saad}], \S\kern1pt5.3.2).
Independently of the value of the quark mass, 
only few iterations then are in general sufficient
to reach the point where 
the rate of convergence of the GCR solver
is not significantly improved by performing further iterations.
This is so in part because $\DB$ tends to be 
well-conditioned, but it should also be noted that
the Schwarz preconditioner is applied when
the accuracy of the current approximate solution 
$\psi_k$ is anyway limited.

The update step (2.8) takes the current residue $\rho=\eta-D\psi$
as input and modifies the field $\psi$ on the block $\B$.
It is thus an operation that acts on these
two globally defined fields in the following way:

\vskip1.0ex
\noindent%
(a)~The restriction of the residue $\rho$ to the block
$\B$ is copied to a block field $\rho_{\B}$. An approximate solution
$\zeta_{\B}$ of the block equation 
$\DB\zeta_{\B}=\rho_{\B}$ is then obtained by applying
a fixed number $\nmr$ of minimal residual iterations.

\vskip1.0ex
\noindent%
(b)~$\psi$ is replaced by $\psi+\zeta_{\B}$ on $\B$
and is left untouched elsewhere. 
The residue $\rho$ is also updated by adding 
$\tilde{\rho}_{\B}=\rho_{\B}-\DB\zeta_{\B}$
on $\B$ and subtracting $\DdBs\zeta_{\B}$
on $\dB$. Note that the minimal residual algorithm
returns both the approximate solution $\zeta_{\B}$ and the associated
residue $\tilde{\rho}_{\B}$.

\vskip1.0ex
\noindent%
Besides $\Msap\eta$ this implementation of the 
Schwarz procedure yields $D\Msap\eta$ essentially for free.
There are two parameters,
the number $\ncy$ of Schwarz cycles and the number $\nmr$ of 
minimal residual iterations performed in step (a),
which should be adjusted so as to minimize the 
average computer time needed to obtain the solution of 
eq.~(2.4) to the desired accuracy.

\subsection 4.2 Even-odd preconditioning

If the Schwarz blocks are chosen to have an even number of points
in all dimensions, even-odd preconditioning may be applied
to accelerate the block solver. 
This simple modification is certainly worth the 
additional programming
effort, since most of the computer
time is spent in step (a).
 
On the full lattice, 
the even-odd preconditioned form of the Wilson--Dirac equation
(2.4) reads
\equation{
   \Dhat\psi_{\rme}=\eta_{\rme}-
   D_{\rme\rmo}(D_{\rmo\rmo})^{-1}\eta_{\rmo},
   \enum
   \next{2ex}
   \Dhat\equiv D_{\rme\rme}-D_{\rme\rmo}(D_{\rmo\rmo})^{-1}D_{\rmo\rme},
   \enum
}
where the subscripts ``e" and ``o" refer to 
the even and the odd lattice sites.
Equation (4.1) determines $\psi_{\rme}$, while
the solution on the odd points,
\equation{
   \psi_{\rmo}=(D_{\rmo\rmo})^{-1}
   \left(\eta_{\rmo}-D_{\rmo\rme}\psi_{\rme}\right),
   \enum
}
is obtained algebraically.

All these formulae carry over to the block equation
$\DB\zeta_{\B}=\rho_{\B}$ without any modifications.
In particular, the even-odd preconditioned
Wilson--Dirac operator~$\Dhat_{\B}$
is again given by eq.~(4.2), with $D$ replaced by $\DB$.
The minimal residual algorithm may then be used to solve the
preconditioned system, in which case
steps (a) and (b) together require
the equivalent of $\nmr+1$ applications
of $\Dhat_{\B}$ (if $\nmr$ iterations are performed)
plus the calculation of about $2\nmr$ scalar products and linear combinations.

\subsection 4.3 Communication overhead

On parallel computers each processor operates on a sublattice that
is usually taken to be a rectangular block similar to the blocks
on which the Schwarz procedure operates. In order to 
minimize the communication
overhead, the local lattice should obviously be 
divisible by the Schwarz blocks.
The block solver does not require any communications in this case,
and it is only in step (b), when 
$\DdBs\zeta_{\B}$ is subtracted from the current residue,
that some data need to be exchanged.
In terms of ``communicated words per arithmetic operation", 
a reduction by a factor $\nmr+1$ is thus achieved
relative to the overhead in the program for the Wilson--Dirac operator.

Another factor of $2$ in the communication overhead 
can actually be saved by noting that
\equation{
   (1-n_{\mu}\dirac{\mu})\DdBs\zeta_{\B}(x)=0
   \quad\hbox{for all}\quad x\in\dB,
   \enum
}
where $n_{\mu}$ is the outward normal vector to the boundary $\dB$
at $x$. 
There are thus two linear relations among the four Dirac components of 
$\DdBs\zeta_{\B}(x)$, and it is therefore 
sufficient to communicate the upper two components in the 
cases where a communication is required.%
\kern1pt\footnote{$\dagger$}{\footnotefont%
It is assumed here that 
the Dirac matrices are in a chiral representation,
where $\dirac{5}=\dirac{0}\dirac{1}\dirac{2}\dirac{3}$ is a 
diagonal matrix with entries $1,1,-1,-1$ on the diagonal. 
The upper two components of the boundary field $\DdBs\zeta_{\B}$
are then guaranteed to be linearly independent.}

Taking this into account, the update of the residue in step (b) proceeds
by first adding $\tilde{\rho}_{\B}$
on all points in the block.
The upper two components of $\DdBs\zeta_{\B}$
are then calculated and stored in an auxiliary array.
Some parts of this field may reside on the exterior boundary of 
the local lattice, in which case they have to be communicated to 
the appropriate neighbouring processes.
Thereafter the full Dirac spinors $\DdBs\zeta_{\B}$ are reconstructed,
using eq.~(4.4),
and subtracted from the residue. Note that the communication
and subtraction can be
delayed until all black (or all white) blocks have been visited.
This allows the relevant parts of the fields on the 
boundaries of all the black (white) blocks
to be transferred in a single data package per direction.

\subsection 4.4 Non-algorithmic accelerations

Since the Schwarz preconditioner does not need to be accurately implemented,
one may just as well use single-precision
arithmetic in the program for the block solver.
The GCR algorithm may also be 
coded in this way, with scalar products accumulated in double precision, 
provided it is restarted before the residue has decreased by 
more than a few orders of magnitude. Evidently, at each
restart of the algorithm, the last residue should be 
recalculated using double-precision arithmetic to
ensure that the solution of the Wilson--Dirac equation is obtained
to full precision (see ref.~[\ref{NumMethods}]
for a more detailed discussion of this point).

All current PC processors support vector instruction sets
({\tt SSE}, {\tt SSE2}, {\tt 3DNow!}~or {\tt AltiVec}) 
that allow short vectors of floating-point numbers to be 
processed simultaneously [\ref{SSE}].
The block solver and other time-critical
parts of the program can be significantly accelerated  
using these additional capabilities, especially so if
the fields on the Schwarz blocks fit into the cache of 
the processors.

\section 5. Tests of the algorithm

As is generally the case for domain decomposition methods, 
the proposed algorithm is expected to perform particularly 
well on parallel computers.
At this point nothing is known, however, about the convergence
properties of the algorithm in the relevant range of parameters.
The tests reported in this section serve to fill this gap
and to show that an excellent parallel efficiency is indeed achieved
on the computer where the tests were carried out
(see appendix B for the technical specifications).

\subsection 5.1 Choice of parameters

Different parameter sets have been considered, but most results
were obtained for a particular choice, defined in the following lines, 
that will be referred to as the default parameter set. 
In general the strategy of the tests is
to distribute the computations 
required for the solution of the Wilson--Dirac equation 
on a fixed lattice over
an increasing number of processors, so that the effects of the 
communication overhead can be clearly seen.
Taking this into account, and the fact that 
the available computer has $64$ processors,
it was decided to perform the tests on a $16^4$ lattice.

Samples of $100$ statistically independent gauge field
configurations were then generated, using the Wilson plaquette action
at (inverse) coupling $\beta=5.9$, where
the lattice spacing is about $0.10\,\fm$.
In each test run, five values of the bare quark mass $m_0$ 
were considered, corresponding to values of $\kappa=(8+2m_0)^{-1}$
equal to $0.1566$, $0.1574$, $0.1583$, $0.1589$ and $0.1592$.
According to a recent large-scale study 
by the CP-PACS collaboration [\ref{CPPACS}],
the ``pion" mass $m_{\pi}$ 
decreases from about $579\,\MeV$ to $320\,\MeV$ in this range of
quark masses.

The sizes of the Schwarz blocks are constrained by
the requirement that they should divide the local lattices.
Larger blocks tend to yield a better preconditioner,
but the block solver then consumes more 
processor time per lattice point.
The exact choice is not critical, and a block size of 
$8\times4^3$ appears to be a good compromise for the given
lattice and machine parameters.
As for the remaining free parameters of the algorithm,
some experimenting suggests to set $\nmr=4$, $\ncy=5$ and
$\nkv=16$, where the latter limits the number of
Krylov vectors that are generated by the GCR recursion before
it is restarted. 

\subsection 5.2 Reference algorithm

To be able to judge the overall performance of the new algorithm, 
timings were also taken for the BiCGstab algorithm [\ref{BiCGstabI}],
which is probably the fastest Krylov space solver for the 
Wilson--Dirac equation [\ref{BiCGstabII}].
The algorithm is applied to the even-odd preconditioned
system (4.1), using a highly optimized code.
In particular, as in the case of the Schwarz procedure 
discussed in subsect.~4.3,
the program only transfers the upper two components of the 
Dirac spinors that need to be communicated,
thus saving
a factor of $2$ in the communication overhead.

\topinsert
%Blanke Zahl
\newdimen\digitwidth
\setbox0=\hbox{\rm 0}
\digitwidth=\wd0
\catcode`@=\active
\def@{\kern\digitwidth}
\tablecaption{Timings in $\mu s$ [speed in Gflop/s]
per lattice point and application of $\Dhat$}
\vskip-1.0ex
$$\vbox{\settabs\+&%
                  xxxxxxxxx&&%      number of processors
                  xxxxxxxxxx&&%      local lattice
                  ixxxxxxxxxxxxxx&&%     M_sap*psi
                  ixxxxxxxxxxxxx&&%     Dhat*psi
                  xxxxxxxxxxxx&&%    SAP+GCR
                  xxxxxxxxx&\cr%     BiCGstab
\thicktablerule
\vskip1.1ex
                \+& \hfill{\ninerm number of}\hfill
                 && \hfill{\ninerm local}\hfill
                 &\cr
\vskip-0.7ex
                \+& \hfill{\ninerm processors}\hfill
                 && \hfill{\ninerm lattice}\hfill
                 &\cr
\vskip-3.8ex
                \+& \hfill
                 && \hfill
                 && \hfill $(\Msap\psi)_{32\,{\rm bit}}$\hfill
                 && \hfill $(\Dhat\psi)_{64\,{\rm bit}}$\hfill
                 && \hfill{\ninerm SAP+GCR}\hfill
                 && \hfill{\ninerm BiCGstab}\hfill
                 &\cr
\vskip1.6ex
\thintablerule
\vskip1.5ex
  \+& \hfill $1$\hfill
  &&  \hskip2.5ex $16^4$\hfill 
  &&  \hfill $0.64\kern4pt[2.35]$\hfill
  &&  \hfill $1.39\kern4pt[0.98]$\hfill
  &&  \hfill $0.75$\hfill 
  &&  \hfill $2.2$\hskip4ex
  &\cr
\vskip0.3ex
  \+& \hfill $2$\hfill
  &&  \hskip2.5ex $16^3\times8$\hfill 
  &&  \hfill $0.86\kern4pt[1.74]$\hfill
  &&  \hfill $2.59\kern4pt[0.52]$\hfill
  &&  \hfill $1.12$\hfill 
  &&  \hfill $4.2$\hskip4ex
  &\cr
\vskip0.3ex
  \+& \hfill $4$\hfill
  &&  \hskip2.5ex $16^2\times8^2$\hfill 
  &&  \hfill $0.86\kern4pt[1.74]$\hfill
  &&  \hfill $3.00\kern4pt[0.45]$\hfill
  &&  \hfill $1.10$\hfill 
  &&  \hfill $4.7$\hskip4ex
  &\cr
\vskip0.3ex
  \+& \hfill $8$\hfill
  &&  \hskip2.5ex $16\times8^3$\hfill 
  &&  \hfill $0.88\kern4pt[1.71]$\hfill
  &&  \hfill $3.42\kern4pt[0.40]$\hfill
  &&  \hfill $1.11$\hfill 
  &&  \hfill $5.2$\hskip4ex
  &\cr
\vskip0.3ex
  \+& \hfill $16$\hfill
  &&  \hskip2.5ex $8^4$\hfill 
  &&  \hfill $0.90\kern4pt[1.65]$\hfill
  &&  \hfill $4.17\kern4pt[0.33]$\hfill
  &&  \hfill $1.13$\hfill 
  &&  \hfill $5.8$\hskip4ex
  &\cr
\vskip0.3ex
  \+& \hfill $32$\hfill
  &&  \hskip2.5ex $8^3\times4$\hfill 
  &&  \hfill $0.94\kern4pt[1.58]$\hfill
  &&  \hfill $4.76\kern4pt[0.28]$\hfill
  &&  \hfill $1.24$\hfill 
  &&  \hfill $6.7$\hskip4ex
  &\cr
\vskip0.3ex
  \+& \hfill $64$\hfill
  &&  \hskip2.5ex $8^2\times4^2$\hfill 
  &&  \hfill $0.98\kern4pt[1.53]$\hfill
  &&  \hfill $5.50\kern4pt[0.25]$\hfill
  &&  \hfill $1.21$\hfill 
  &&  \hfill $7.3$\hskip4ex
  &\cr
\vskip1.0ex
\thicktablerule
}
$$
\vskip-2ex
\endinsert

A technical detail worth mentioning here is that 
single-precision acceleration [\ref{NumMethods}]
does not work too well in this case, because
the increased arithmetic speed is practically compensated by
the fact that more iterations are required 
to reach a specified precision.
Moreover, there is a negative effect on the stability of 
the al\-go\-rithm. The 
BiCGstab solver was therefore implemented in double-precision
arith\-metic,
with all time-critical parts programmed using {\tt SSE2} vector instructions.

\subsection 5.3 Basic program and algorithm performance 

In terms of numbers of floating-point operations, 
the Schwarz preconditioner $\Msap$ is roughly equivalent to
$(\nmr+1)\kern1pt\ncy$ applications of $\Dhat$.
The speed of the programs that implement the preconditioner
and the complete solver
(denoted ``SAP+GCR" in the following) is therefore best 
measured in units of execution time per local lattice point
and equivalent application of $\Dhat$. 

\topinsert
\vbox{
\vskip0.0cm
\epsfxsize=10.0cm\hskip1.0cm\epsfbox{perf1.eps}
\vskip0.2cm
\figurecaption{%
Average 
execution time $t$ needed to 
solve the Wilson--Dirac equation (2.4)
as a function of the bare quark mass. A single processor is
used here and the solvers are stopped 
when $\|\eta-D\psi\|<10^{-8}\|\eta\|$. The lattice and algorithmic
parameters are as specified in subsect.~5.1, and the dotted lines 
are drawn to guide the eye.
}
\vskip0.0cm
}
\endinsert

For the case where a single processor is used,
some relevant performance figures 
are quoted in the first line of table~1.
The associated Gflop/s rates, given in square brackets, 
refer to the actual number of floating-point 
operations. It should again be emphasized at this point
that the use of single-precision arithmetic for the 
Schwarz preconditioner is entirely adequate and does not
set a limit on the precision that can be reached by 
the SAP+GCR solver.
Note that a BiCGstab iteration involves two applications of $\Dhat$,
so that the time per iteration and lattice point
is twice the figure quoted in table~1. 
The linear algebra in the BiCGstab algorithm
thus consumes more than a third of the total time
spent by the single-processor program.
This overhead is much smaller 
in the case of the SAP+GCR solver, because 
the application of the Schwarz preconditioner
is relatively expensive in terms of computer time.

\topinsert
\vbox{
\vskip0.0cm
\epsfxsize=7.0cm\hskip2.5cm\epsfbox{pareff.eps}
\vskip0.2cm
\figurecaption{%
Execution time required for the solution of the Wilson--Dirac equation
(2.4) as a function of the number $n_p=2,4,\ldots,64$ 
of processors used by the program.
The parameters are as in table~1 and fig.~4, with the quark mass
fixed to the smallest value considered there. 
Ideal parallel scaling is represented by the dashed lines.
}
\vskip0.0cm
}
\endinsert

In fig.~4 the
time needed for the solution of the Wilson--Dirac equation
to a speci\-fied precision is plotted versus $1/\mq$, where 
$\mq=m_0-\mc$ denotes the additively renormalized quark mass
and $\mc$ the critical bare mass. 
The SAP+GCR solver thus obtains
the solution faster than the BiCGstab algorithm, by a factor
ranging from $1.6$ to $1.9$, if a single processor is used.
When going from large to small quark masses, 
the curvature in the data suggests that the BiCGstab solver becomes
relatively more efficient,
but this trend is less pronounced on larger lattices 
and should therefore be taken as a finite-volume
anomaly
(cf.~subsect.~5.5). 

Another result of these tests is that 
the GCR algorithm converges in comparatively few steps.
On average only $17$ Krylov vectors are generated 
at the largest quark mass and $64$ at the smallest mass, while
$140$ to $437$ BiCGstab iterations are needed
to obtain the solution to the specified precision.
Since both algorithms have similar theoretical
convergence properties [\ref{Saad}], this suggests that the condition
number of the Schwarz 
preconditioned Wilson--Dirac operator $D\Msap$ is roughly
$15$ times smaller than the condition number of $\Dhat$, 
at all quark masses that have been considered.

\subsection 5.4 Parallel efficiency

Evidently these algorithmic properties are unchanged when the 
computational work is distributed over several processors,
and the variations in the performance figures listed in table~1
thus provide a direct measure for the parallelization 
losses.
The lattice and algorithmic parameters 
have been set to the default values 
in all these tests.
In particular, 
the dimensionality
of the parallelization and the communication overhead
are increasing as one moves down the table. 

\topinsert
\vbox{
\vskip0.0cm
\epsfxsize=10.0cm\hskip1.0cm\epsfbox{perf16.eps}
\vskip0.2cm
\figurecaption{%
Same as fig.~4, but now for a $48\times24^3$ lattice distributed
over $16$ processors. The sizes of the local lattices and the Schwarz blocks
are $24\times12^3$ and $6^2\times4^2$ respectively, and 
$\nkv$ has been set to $24$ in this case. 
All other parameters are unchanged.
}
\vskip0.0cm
}
\endinsert

When going from $1$ to $2$ processors,
the steep rise in the timings is, however, only partially 
due to the communication overhead.
The principal cause for the slow-down instead is 
the fact that the two processors on the nodes of the computer 
share the available memory bandwidth (cf.~appendix B).
For this reason the parallel efficiency of the programs should be 
judged by comparing the many-processor timings
with the performance of the two-processor programs rather than 
the single-processor ones.

The results quoted in table~1 show that the parallel 
efficiency of the programs for 
the Schwarz preconditioner and the SAP+GCR algorithm
is very good.
If the lattice is distributed over all $64$ processors,
for example, the surface-to-volume ratio of the local lattices
is as large as $1.5$, but 
the parallelization overhead 
in the program for the preconditioner nevertheless 
only consumes about $12\%$
of the execution time.
Moreover, this figure is even smaller in the case of the solver, because
practically no communications are required in the linear algebra programs.
Note incidentally that the computer achieves a total sustained speed of 
nearly $100$ Gflop/s at this point.

Given the available network bandwidth, it is no surprise, however,
that the program for the even-odd preconditioned lattice Dirac operator
and the BiCGstab solver do not scale as well.
The difference is perhaps made clearer by  
fig.~5, where the time needed to solve the Wilson--Dirac equation
is plotted as a function of the number of processors.
In particular, if all $64$ processors are used,
and depending on the quark mass,
the SAP+GCR solver is from $3.3$ to
$4.1$ times faster than the BiCGstab solver.

\subsection 5.5 Larger lattices

The timings taken on larger lattices
suggest that the performance pattern remains practically the same
as the one seen so far.
Essentially
the communication losses are
determined by the surface-to-volume ratio of the local lattices,
while the number of processors that are used
appears to be less relevant.

For illustration, the time needed to solve the Wilson--Dirac equation
on a $48\times24^3$ lattice is plotted in fig.~6. In this
test $16$ processors were used and the surface-to-volume ratio
of the local lattices was only $0.6$. The plot shows that
the SAP+GCR and the BiCGstab algorithms 
have a similar scaling behaviour
as a function of the quark mass, the SAP+GCR solver being faster by
a factor ranging from $2.5$ to $2.9$. 
Moreover, the ratio of the iteration counts of the two algorithms is not very
different from the one observed before,
although the convergence is generally slower on this lattice
than on the $16^4$ lattice.

\section 6. Conclusions

The algorithm presented in this paper provides the
first example for a successful application of 
a domain decomposition method in lattice QCD.
Traditionally the lattice Dirac equation is solved
using a general purpose algorithm such as the BiCGstab solver.
Domain decomposition methods are fundamentally different from these,
because they do not operate in the usual Krylov spaces.
The locality and ellipticity of the equation
are instead exploited to construct a highly effective preconditioner.

An obvious advantage of domain decomposition methods is that 
they are usually
well suited for parallel processing.
In the case of the algorithm
described here, for example, 
the number of words per arithmetic operation
which must be communicated is smaller than
the communication rate in the BiCGstab algorithm
by about a factor $5$. 
This may not be an impressive figure,
but it should be noted in this connection that
a reduction of the communication overhead
by a factor $2$ alone allows $16$ times more processors
to be used efficiently once the program is parallelized in four dimensions
(and runs well in this mode).

In terms of condition numbers, 
the non-overlapping Schwarz procedure
that was considered in this paper
yields an excellent preconditioner 
for the Wilson--Dirac operator. 
The numerical tests showed this very clearly,
but another outcome of the tests was that 
the scaling behaviour of the SAP+GCR solver
with respect to the quark mass is rather similar to
the one of the BiCGstab algorithm. It is conceivable
that more sophisticated domain decomposition methods
can do better than this, and 
there are in fact quite a few cases in other branches of science,
where such methods come close to having an ideal (i.e.~constant) scaling 
behaviour [\ref{QuarteroniValli}]. For lattice QCD any progress
in this direction would evidently be a major step forward.

\vskip1ex
The numerical work reported in this paper 
was carried out on a PC-cluster at 
the Institut f\"ur Theo\-retische Physik der Universit\"at Bern,
which was funded in part by 
the Schweizerischer Nationalfonds.
I wish to thank both institutions for the support
given to this project and Peter Weisz for a critical reading 
of the manuscript.

\vskip1ex

%\vfill\eject

\appendix A. Programming issues

The remarks collected in this appendix
provide some guidelines for the programming of the Schwarz procedure.
They are based on the experience gained in the course of this work,
where all programs were written in C, using 
the standard message passing interface (MPI) for the communications
between the processors. Evidently, the recommended strategies may not
be appropriate in other programming environments.

\subsection A.1 Object orientation

The Schwarz procedure is an obvious case for
an object-oriented approach.
A block of lattice points, 
for example, is best represented by a structure that 
contains the block fields and the local geometry arrays.
There should be enough predefined information in this structure
that the block solver can proceed without reference to 
the surrounding lattice. In particular, it is advantageous
to have a copy of the relevant gauge field variables on each block.
Other useful objects are block boundaries and 
the block grids on which the Schwarz procedure operates.

Among the 
members of the block and boundary structures,
the geometry arrays play a central r\^ole. 
Lattice points are usually labelled by an integer
index {\tt ix} in an arbitrary way.
The embedding in the global lattice of a block point 
with local index {\tt ix} thus amounts to specifying
its global index {\tt imb[ix]}.
Similarly each boundary
point with local index {\tt ix} has a unique partner point
in the block with block index {\tt ipp[ix]}.
This approach views
blocks, boundaries and the global lattice
as nothing more than ordered sets of points,
whose geometry is encoded in the 
appropriate index arrays.

Once the structures representing the
objects have been defined, some
utility programs need to be
written for object creation, destruction and other basic manipulations.
In particular, since blocks have their own fields, there must be functions
that copy the relevant parts of the global fields to 
and from a specified block.

\subsection A.2 Generic programs

The programming effort as well as the program sizes
can be significantly reduced by writing generic programs.
Linear algebra routines, for example,
only need to know the starting addresses of the spinor fields
on which they operate and the number of spinors in the fields.
The same programs can then be applied independently of 
the geometric context.

In the case of the programs for the Wilson--Dirac operator,
a generic code requires as input
the addresses of the gauge field, of the source and 
the target spinor fields, and of the nearest-neighbour index array.
Evidently
the quark mass and the number of elements in the fields must also be
specified, but apart from this the program does not need any further
information to be able to proceed.
Different local geometries and boundary conditions
are then easily taken into account through predefined index arrays.

\subsection A.3 Parallelization

If the global lattice is to be distributed over a large number 
of processors,
it will be 
preferable, in general, to parallelize the program 
in four dimensions.
Communications in eight directions are required in this case and 
some care should obviously be taken to ensure that the associated overhead
remains small.

Data transfers via MPI send and receive functions are 
characterized by the point-to-point bandwidth
of the network and the start-up latency. The latter 
can be kept to a minimum by
using so-called persistent communication requests and by
aligning the data arrays in memory so that any
unnecessary data packing is avoided.
It may actually be quite important to follow this advice,
since the data packages that are exchanged in the course
of the Schwarz procedure tend to be relatively small.
In particular, advantage should be taken of 
the chessboard colouring of the Schwarz blocks
to maximize the package sizes, as explained at the 
end of subsect.~4.3.

\appendix B. Machine parameters

The tests reported in sect.~5 were performed
on a commodity PC-cluster with $32$ nodes,
each consisting of a motherboard with 
Intel E7500 chipset, 
two Intel Xeon processors ($2.4$ GHz, $512$ KB cache),
$1$ GB of shared ECC DDR-200 memory and a Myrinet-2000 communication card.
Fibre cables connect the Myrinet cards to a fast switch,
which allows for arbitrary 
bidirectional point-to-point communications between the 
nodes.

The tested programs are written in C
with all communications (including intra-node data transfers) 
programmed using standard MPI functions.
In most cases the best performance is obtained by having
one process on each processor. In particular,
if the program is written so that the two processors on the
nodes send and receive data in an 
alternating fashion, the bidirectional bandwidth
of the Myrinet network can then be saturated without any further 
programming effort.

On the software side, the gcc compiler version 3.2.2 was used
with optimization options {\tt -mcpu=i586 -malign-double -fno-force-mem -O}.
Following ref.~[\ref{SSE}], {\tt SSE} and
{\tt SSE2} vector instructions were included through
embedded assembler statements.
The programs were then linked to 
the MPICH-over-GM library provided by Myricom
to enable the MPI functionality.

In linear algebra programs (where the time required for the arithmetic 
operations is negligible), the effective total memory-to-processor bandwidth
on each node
is in the range from $1.6$ to $2.4$ GB/s.
For bidirectional node-to-node communications and 
data packages larger than $1$ MB, the network
achieves a throughput of $160$ MB/s in each direction.
If persistent communication requests are used (in MPI parlance),
the effective bandwidth per link and direction is still $92$ MB/s for packages
as small as $4$ KB. Unidirectional data transfers are 
generally faster and reach about $240$ MB/s.

\beginbibliography

% Domain decomposition

\bibitem{QuarteroniValli}
A. Quarteroni, A. Valli,
Domain decomposition methods for partial differential equations
(Oxford University Press, Oxford, 1999)

\bibitem{Saad}
Y. Saad, Iterative methods for sparse linear systems,
2nd ed. (SIAM, Philadelphia, 2003); see also
{\tt http://www-users.cs.umn.edu/\~{}saad/}

% Application of the SAP to lattice QCD

\bibitem{LuscherSchwarz}
M. L\"uscher,
JHEP 0305 (2003) 052

% Schwarz alternating procedure

\bibitem{Schwarz}
H. A. Schwarz, Gesammelte Mathematische Abhandlungen, vol. 2
(Springer Verlag, Berlin, 1890)

% O(a) improved lattice QCD

\bibitem{Wilson}
K. G. Wilson, Phys. Rev. D10 (1974) 2445

\bibitem{SW}
B. Sheikholeslami, R. Wohlert,
Nucl. Phys. B 259 (1985) 572

\bibitem{OaImp}
M. L\"uscher, S. Sint, R. Sommer, P. Weisz,
Nucl. Phys. B478 (1996) 365

% GMRESR

\bibitem{GMRESRI}
H. A. van der Vorst, C. Vuik,
Num. Lin. Alg. Appl. 1 (1994) 369

\bibitem{GMRESRII}
C. Vuik,
J. Comput. Appl. Math. 61 (1995) 189

% Low-mode preconditioning

\bibitem{NumMethods}
L. Giusti, C. Hoelbling, M. L\"uscher, H. Wittig,
Comput. Phys. Commun. 153 (2003) 31

% Use of SSE/SSE2 instructions

\bibitem{SSE}
M. L\"uscher,
Nucl. Phys. B (Proc. Suppl.) 106 (2002) 21

% The ultimate quenched calculation with Wilson quarks

\bibitem{CPPACS}
S. Aoki et al. (CP-PACS collab.),
Phys. Rev. D67 (2003) 034503

% BiCGstab inverter

\bibitem{BiCGstabI}
H. A. van der Vorst, 
SIAM J. Sci. Stat. Comput. 13 (1992) 631

\bibitem{BiCGstabII}
A. Frommer, V. Hannemann, B. N\"ockel, T. Lippert, K. Schilling,
Int. J. Mod. Phys. C5 (1994) 1073

\endbibliography

\bye